\documentclass[prl,twocolumn,amssymb,amsmath,floatfix]{revtex4}

\usepackage{graphicx}
\usepackage{dcolumn}
\usepackage{bm}
\usepackage{amsmath}
\usepackage{chemformula}
\usepackage{booktabs}
\usepackage{multirow}
\usepackage{newtxmath}
\usepackage{times}
\usepackage{amssymb}
\usepackage{latexsym}
\usepackage{epsfig}
\usepackage{amsbsy}
\usepackage{array}
\usepackage{amssymb}
\usepackage{setspace}
\usepackage{bm}
\usepackage{tabularx}
\usepackage{color}

\def\sint{\ifmmode{- \!\!\!\!\!\! \int}
    \else{\hbox{$- \!\!\!\! \int \ $}}\fi}

\begin{document}

\preprint{Physical Review Letters}

\title{Predictive model of surface adsorption in dissolution on transition metals and alloys}
\author{Bo Li}
\author{Wang Gao}
\email{wgao@jlu.edu.cn}
\author{Qing Jiang}
\date{\today}
\affiliation{Key Laboratory of Automobile Materials (Jilin University), Ministry of Education, Department of Materials Science and Engineering, Jilin University, Changchun 130022, China}

\begin{abstract}

Surface adsorption, which is often coupled with surface dissolution, is generally unpredictable on alloys due to the complicated alloying and dissolution effects. Herein, we introduce the electronic gradient and cohesive properties of surface sites to characterize the effects of alloying and dissolution. This enables us to build a predictive model for the quantitative determination of the adsorption energy in dissolution, which holds well for transition metals, near-surface alloys, binary alloys, and high-entropy alloys. Furthermore, this model uncovers a synergistic mechanism between the $d$-band upper-edge ratio, $d$-band width and $s$-band depth in determining the alloying and dissolution effects on adsorption. Our study not only provides fundamental mechanistic insights into surface adsorption on alloys but also offers a long-sought tool for the design of advanced alloy catalysts.

\end{abstract}


\maketitle

Alloys, covering from near-surface alloys (NSAs), binary alloys (BAs), to high-entropy alloys (HEAs), are essential for heterogeneous catalysis and anticorrosion engineering \cite{greeley2004alloy,stamenkovic2007trends,yeh2004nanostructured,ma2017first}. Their catalytic activity, which is governed by the adsorption energy in terms of the Sabatier principle \cite{sabatier1913catalyse} and Brønsted–Evans–Polanyi (BEP) relation \cite{bronsted1928acid,evans1938inertia,bligaard2004bronsted}, is the footing stone for the development of electrode materials. However, when the adsorbates are continually adsorbed to the surfaces of metallics during the catalytic process, the surfaces also suffer severe corrosion under electrochemical condition with the surface-atoms dissolution \cite{wang2011multimetallic,topalov2012dissolution,kang2014multimetallic,lopes2016relationships,lopes2020eliminating}. Namely, the adsorption of surface sites is generally competitive and coupled with surface dissolution, prohibiting the establishment of principles for guiding material design. 

The main challenge stems from the alloying and dissolution effects, since alloying and dissolution can generate a diverse coordination environment of surface sites. In particular, the random composition and chemical ordering of HEAs offer multiple active centers, possess significant charge redistribution, and create different crystal phases, all of which bring more opportunities to modulate the electronic and geometric structures of surface, unfortunately also making it more difficult to explicitly understand the picture of alloying effects \cite{loffler2018discovery,xie2019highly,batchelor2019high,lu2020neural,xin2020high,yao2020computationally}. Therefore, the current understanding of adsorption achieves mainly on simple metallics instead of complex alloys \cite{hammer1995electronic,hammer1995gold,hammer2000theoretical,abild2007scaling,vojvodic2009electronic,vojvodic2014electronic,xin2014effects,inouglu2010simple,mpourmpakis2010identification,calle2014fast,calle2015finding,ma2017orbitalwise,calle2012physical,gao2020determining,li2020electronic}. The adsorption energy has been correlated to the \textit{d}-band center or upper edge in \textit{d}-band model \cite{hammer1995electronic,hammer1995gold,hammer2000theoretical,vojvodic2009electronic,vojvodic2014electronic,xin2014effects,inouglu2010simple}, coordination number \cite{mpourmpakis2010identification,calle2014fast,calle2015finding,ma2017orbitalwise}, valence and electronegativity of surface atoms \cite{calle2012physical,gao2020determining,li2020electronic}, which hold for transition metals (TMs), TM nanoparticles (NPs), and partial NSAs. However, it still presents a fundamental challenge to uncover the mechanism and do the quantitative prediction for the adsorption on various alloys (e.g. BAs and HEAs), especially in dissolution.

Herein we introduce the electronic gradient and cohesive properties of surface sites to characterize the effects of alloying and dissolution. Accordingly, a predictive model has been established to accurately determine the adsorption energy in dissolution on TMs, NSAs, BAs, and HEAs. This model further unravels the electronic origin of metallics in determining the adsorption energy, which outlines a novel physical picture for understanding surface adsorption on alloys.

To ensure the generality of our scheme, we have considered the widespread adsorption of $\rm {CH}_\textit{x}$ (\textit{x} = 0-3), $\rm {CCH_3}$, CO, $\rm {NH}_\textit{x}$ (\textit{x} = 0-2), OH, F and Cl on (100), (110), (111), (211) and (532) surfaces of TMs, NSAs, BAs, and HEAs (Fig. S1 \cite{li2021submission}). The adsorption process in dissolution is illustrated in Note S1, the inset of Fig. 1a and b, and Figs. S2-S4 \cite{li2021submission}. All calculations were performed with Perdew-Burke-Ernzerhof (PBE) \cite{perdew1996generalized} and Perdew-Wang-91 (PW91) \cite{perdew1992atoms} exchange-correlation (XC) functionals (see more details in Note S2 \cite{li2021submission}).
\begin{figure*}
	\centering
	\includegraphics[width=12.5cm]{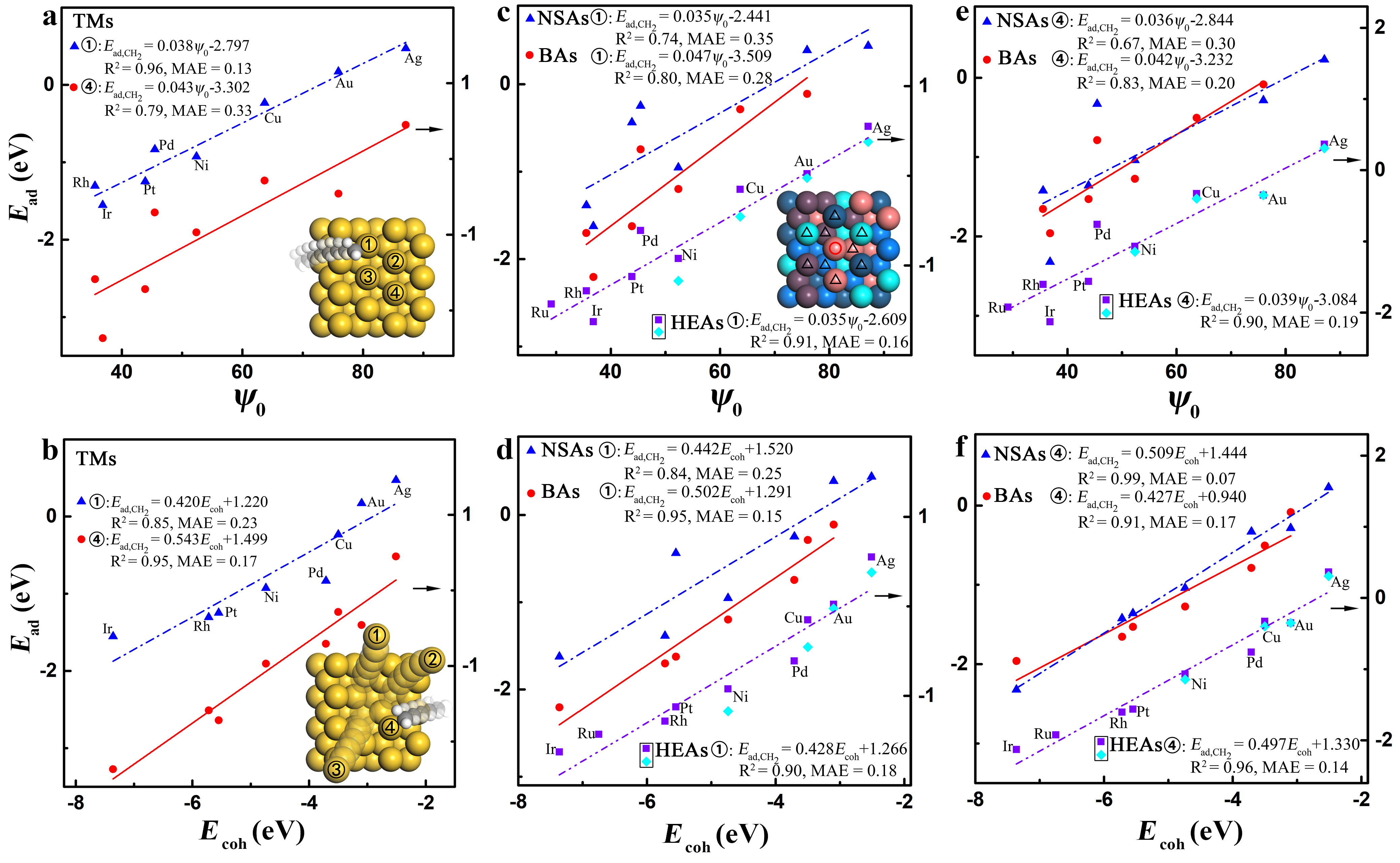}
	\caption{Comparison between the electronic descriptor $\psi_0$ and cohesive energy $E_{\rm {coh}}$ \cite{zhang2014understanding} of adsorption-site atoms in describing the top-site adsorption energy in dissolution. (a) and (b) (111) surface of TMs. (c)-(f) (111) surface of Cu-based NSAs, (100) surface of Ag-based BAs, and (111) surface of RuRhIrPdPt-based HEAs. The symbol numbers \small{\textcircled{\small{1}}}-\small{\textcircled{\small{4}}} of the inset of (a) and (b) denote the adsorption order in dissolution (see details in Figs. S2-S4 and Note S1 \cite{li2021submission}). Taking HEAs as example, the red circle and black triangle of the inset of (c) denote the adsorption site and its nearest neighbors. The violet square and cyan rhombus icons for HEAs in (c)-(f) represent the cases that changing only the element of adsorption sites and that changing all the element that is the same as the adsorption site. }
	\label{fm1}
\end{figure*} 
\hspace{10pt}

To understand the alloying effects on adsorption, we study the active center of adsorption, which includes the adsorption site and its nearest neighbors (see the inset of Fig. 1c and Fig. S1 \cite{li2021submission}). The alloying effects can be roughly separated into the adsorption-site effect and the environment effect. The former refers to the change of chemical composition of adsorption sites with the fixed surrounding properties, whereas the latter does the change of the chemical properties of the nearest neighbors with the fixed adsorption-site properties. We adopt ${\mathop{{{ \left( {{\mathop{ \prod }\nolimits^{{N}}_{{i=1}}{\mathop{{S}}\nolimits_{{\rm {v}\textit{i}}}}}} \right) }}}\nolimits^{{2/N}}/\mathop{{{ \left( {{\mathop{ \prod }\nolimits^{{N}}_{{i=1}}{\mathop{{ \chi }}\nolimits_{{i}}}}} \right) }}}\nolimits^{{1/N}}}$ to calculate the electronic descriptor for the adsorption site $\psi_{{0}}$ and for the active center $\psi_{{1}}$, which correspond to the local and non-local electronic effects respectively. $N$ is the atom number of the adsorption site or the active center, while $S_{\rm{v}\textit{i}}$ and $\chi_i$ represent the valence-electron number (including that of both $d$- and $s$-electrons) and Pauling electronegativity of ${i}$th involved atom. Hence, we introduce the electronic gradient descriptor of surface sites as,
\begin{eqnarray}
	\psi^{'}=\psi_0^m/\psi_1^n 
\end{eqnarray}
\noindent With the change of \textit{m} and \textit{n}, $\psi^{'}$ provides a simple way to estimate the gradient of surface electronic properties of alloys. Notably, $\psi^{'}$ can (partially) characterize the adsorption-site effect or environment effect of alloying, or the coupling of them depending on the surface sites as well as \textit{m} and \textit{n}.  

Now we try to understand the adsorption during the dissolution process. Fig. 1a and b and Fig. S5c and d \cite{li2021submission} show the adsorption energy of \ch{CH2} on TM(111) surface in dissolution. Clearly, the adsorption-site descriptor $\psi_0$ performs well in characterizing the trends of adsorption energy on the undissolved surfaces but gradually loses efficiency in the characterization as the dissolution proceeds. In contrast, the cohesive energy of adsorption-site atoms $E_{\rm {coh}}$ \cite{zhang2014understanding} becomes increasingly accurate in describing the adsorption energy with the dissolution process going on. Based on our findings on the undissolved surfaces of TMs \cite{gao2020determining}, we propose an expression of the adsorption energy in dissolution as,
\begin{eqnarray}
	\begin{aligned}
	E_{\rm{ad}}&=\mu_1\times{D_{\rm{ad}}}+\mu_2\times\overline{CN}+\theta\\
	&=\frac{{1}}{{10}} \times  \frac{{X_{\rm {m}}}-X}{{X_{\rm {m}}}+1}\times \mathop{{D}}\nolimits_{{\rm {ad}}}+\frac{{1}}{{5}}\times \frac{{X+1}}{{\mathop{{X}}\nolimits_{{\rm {m}}}+1}}\times \overline{CN} +\theta 
    \end{aligned}
\end{eqnarray} 
\noindent where $X_{\rm{m}}$ and $X$ correspond to the maximum bondable and actual bonding numbers of the central atom for a given adsorbate. The generalized coordination number $\overline{CN}$ is obtained by dividing the summation of the usual coordination number $CN$ of the adsorption sites$'$ nearest neighbors with the $CN$ in bulk \cite{calle2014fast,calle2015finding}. $\theta$ is a constant for a given adsorbate and is likely controlled by the coupling between the \textit{sp} states of substrates and the states of adsorbates on TMs ($\theta$ can be easily determined by only doing calculations on one substrate) \cite{gao2020determining}. The electronic descriptor ${\rm \textit{D}_{ad}}$ for the adsorption energy on TMs in dissolution can be expressed as,
\begin{eqnarray}
		\begin{aligned}
		&{{D_{\rm{ad}} = k_1\psi^{'}+k_2}}{\left( {\frac{{25}}{{2}}\mathop{{E}}\nolimits_{{\rm {coh}}}+115} \right)}\\ 
		&={\left(\frac{\overline{CN}+1}{CN_{\rm{u}}+1}\right)}^2\psi_0+\left[{1-{\left(\frac{\overline{CN}+1}{CN_{\rm{u}}+1}\right)}^2}\right]\left(\frac{25}{2}E_{\rm{coh}}+115\right)
		\end{aligned}
\end{eqnarray}
\begin{figure*}
	\centering
	\includegraphics[width=12.5cm]{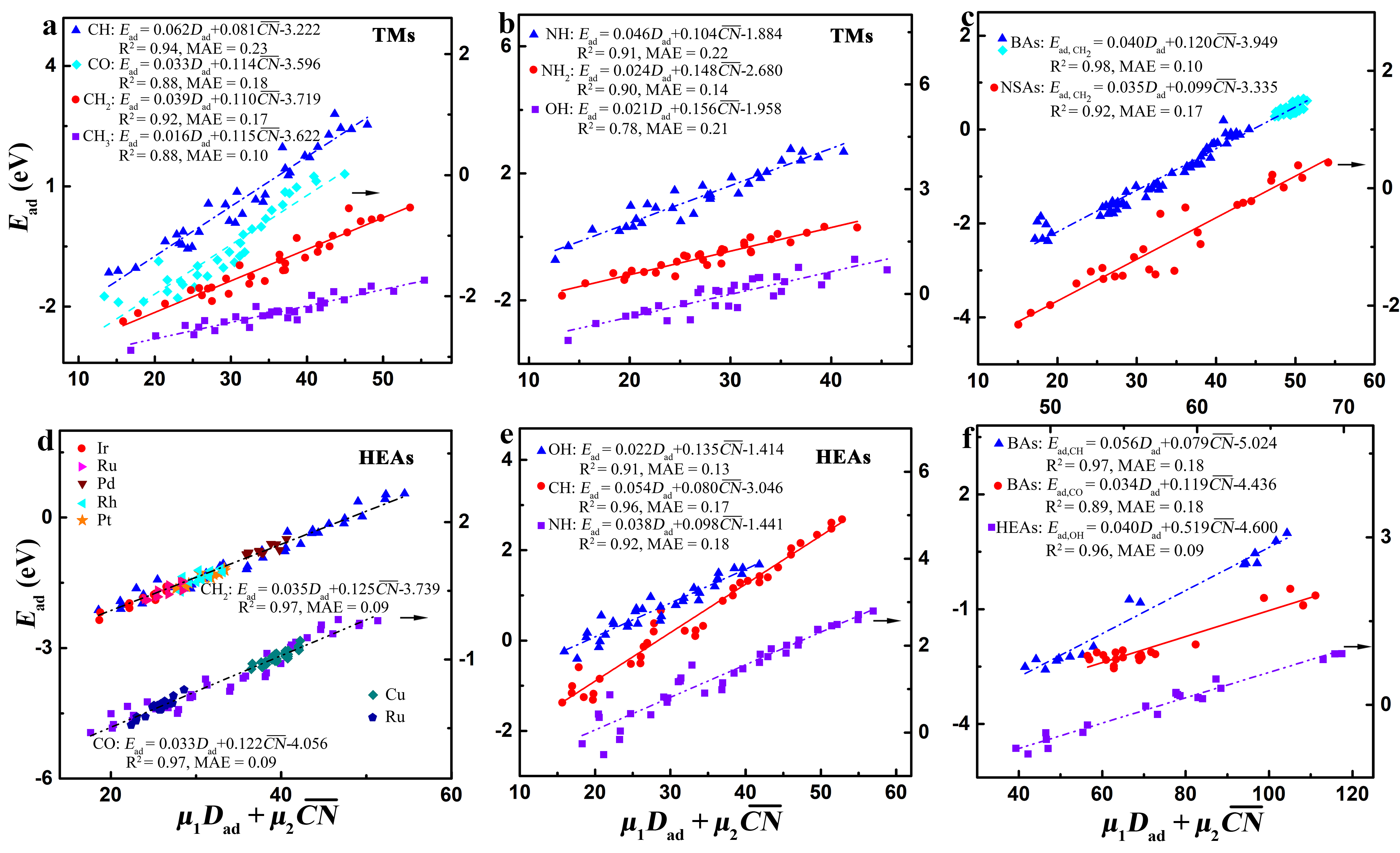}
	\caption{Adsorption energies of the different adsorbates as a function of the proposed descriptors in dissolution. (a) and (b) $\rm {CH}_\textit{x}$ (\textit{x} = 1-3), CO, $\rm {NH}_\textit{x}$ (\textit{x} = 1, 2), and OH on (111) surface of TMs \cite{roling2018structure}. (c) $\rm {CH}_2$ on (100) and (111) surfaces of Ag-based BAs in the adsorption-site effect of alloying (blue triangle) and the environment effect of alloying (cyan rhombus), and (111) surface of Cu-based NSAs in the adsorption-site effect of alloying. (d) $\rm {CH}_2$ and CO on (100), (110), (111), (211) and (532) surfaces of RuRhIrPdPt-based HEAs. The blue triangle and violet square represent the adsorption-site effect of alloying, while the rest do the environment effect of alloying. (e) CH, NH and OH on (111) surface of RuRhIrPdPt-based HEAs in the adsorption-site effect of alloying. The data in (a)-(e) are obtained at the top site. (f) CH at the hcp site and CO at the bridge site on the undissolved (211) surface of AgAu, AgPd, IrRu, and PtRh BAs \cite{andersen2019beyond}, and OH at the bridge site of (111) surface of RuRhIrPdPt-based HEAs in both the adsorption-site and environment effects of alloying.}
	\label{fm1}
\end{figure*}
\begin{table}[]
	\caption{Comparison between the predicted and DFT-calculated prefactors $\mu_1 = 1/10\times(X_{\rm{m}}-X)/(X_{\rm{m}}+1)$ and $\mu_2 = 1/5\times(X+1)/(X_{\rm{m}}+1)$ of Eq. (2) on TMs and HEAs. The calculated values for $\rm {CH_2}$ adsorption are $\mu_1$ = 0.035 and $\mu_2$ = 0.099 on NSAs and $\mu_1$ = 0.040 and $\mu_2$ = 0.120 on BAs. All data are obtained at the top site.}
	\begin{tabularx}{8.6cm}{Xp{0.6cm}<{\centering}p{0.6cm}<{\centering}p{0.8cm}<{\centering}p{0.8cm}<{\centering}p{0.05cm}<{\centering}p{0.8cm}<{\centering}p{0.8cm}<{\centering}p{0.05cm}<{\centering}p{0.8cm}<{\centering}p{0.8cm}<{\centering}}
		\hline \hline
		\multirow{3}{*}{species} &\multirow{3}{*}{$X_{\rm{m}}$} &\multirow{3}{*}{$X$} &\multicolumn{2}{c}{Predicted} & &\multicolumn{5}{c}{DFT-calculated} \\ 
		\cline{4-5}  \cline{7-11} 
		&  &  &\multicolumn{2}{c}{} & &\multicolumn{2}{c}{TMs} & &\multicolumn{2}{c}{HEAs} \\
		\cline{4-5}  \cline{7-8} \cline{10-11}
		&   &   &$\mu_1$  &$\mu_2$         &   &$\mu_1$             &$\mu_2$        &  &$\mu_1$          &$\mu_2$                 \\ 
		\hline
		CH             &4   &1          &0.060         &0.080         &         &0.062      &0.081         &         &0.054       &0.080      \\
		$\rm{CH_2}$    &4   &2       &0.040         &0.120         &         &0.039      &0.110         &         &0.035       &0.125      \\
		$\rm{CH_3}$    &4   &3      &0.020         &0.160         &         &0.016      &0.115         &         &0.015       &0.120      \\
		CO             &4   &2       &0.040         &0.120         &         &0.033      &0.114         &         &0.033       &0.122      \\
		$\rm{CCH_3}$   &4   &1       &0.060         &0.080         &         &0.064      &0.079         &         &            &           \\
		NH             &3   &1       &0.050         &0.100         &         &0.046      &0.104         &         &0.038       &0.098      \\
		$\rm{NH_2}$    &3   &2       &0.025         &0.150         &         &0.024      &0.148         &         &0.025       &0.145      \\
		OH             &2   &1       &0.033         &0.134         &         &0.021      &0.156         &         &0.022       &0.135      \\ 
		\hline	\hline
	\end{tabularx}
\end{table}
\noindent where ${\rm \textit{CN}_{u}}$ is the usual coordination number of adsorption sites on the undissolved surfaces. The coefficients 25/2 and 115 of the $E_{\rm {coh}}$ term stem from the slope and offset of the fitting relation between $\psi_{{{0}}}$ and $E_{\rm {coh}}$ (Fig. S6 \cite{li2021submission}). As the dissolution proceeds, there exists a gradual transition from $\psi_0$ to $E_{\rm {coh}}$ for the electronic determinant of adsorption energy on TMs (Figs. S5, S7-S11, S12a and S12b \cite{li2021submission}).  Eqs. (2) and (3) describe well the adsorption energy of $\rm {CH}_\textit{x}$ (\textit{x} = 1-3), $\rm {CCH_3}$, CO, $\rm {NH}_\textit{x}$ (\textit{x} = 1, 2), and OH at the different adsorption coverage and sites on TMs in dissolution (Fig. 2a and b and Figs. S13-S15 and Note S3 \cite{li2021submission}) \cite{roling2018structure,andersen2019beyond,hoffmann2016framework}, with the fitted mean absolute error (MAE) of 0.16 eV. The fitted prefactors $\mu_1$ and $\mu_2$ are also in good agreement with the predictions by Eq. (2) (Tables I and S1 \cite{li2021submission}). Note that the prefactors $\mu_1$ and $\mu_2$ of Eq. (2), determined by the bonding characters of adsorbates $(X_{\rm{m}}-X)/(X_{\rm{m}}+1)$ and $(X+1)/(X_{\rm{m}}+1)$, can be deduced and rationalized from the first-order approximation of effective medium theory (EMT) and bond-order conservation framework (Note S4 \cite{li2021submission}) \cite{norskov1980effective,gao2020determining}.

We now quantify the adsorption-site effect of alloying on the adsorption energy during dissolution. Fig. 1c-f and Figs. S12c, S12d and S16-S19 \cite{li2021submission} show that $E_{\rm {coh}}$ of adsorption-site atoms generally performs better than the adsorption-site descriptor $\psi_0$ in describing the adsorption energy of $\rm {CH}_\textit{x}$ (\textit{x} = 1-3), CO, $\rm {NH}_\textit{x}$ (\textit{x} = 1, 2), and OH at the top site of NSAs, BAs, and HEAs in the adsorption-site effect of alloying for each dissolution step. For HEAs, changing only the element of adsorption sites or changing all the element that is the same as the adsorption site follows the same linear relation (Fig. 1c-f). By further including the coordination-chemistry effect with the electronic gradient descriptor $\psi^{'}$, we introduce a universal descriptor to quantitatively describe the electronic contribution of alloys to the adsorption energy in dissolution as, 
\begin{eqnarray}
		\begin{aligned}
			D_{\rm{ad}}&=k_1\psi^{'}+k_2\left(\frac{25}{2}E_{\rm{coh}}+115\right)\\
			&=k_1\frac{\psi_0^2}{\psi_1}+k_2\left(\frac{25}{2}E_{\rm{coh}}+115\right)
		\end{aligned}			
\end{eqnarray}
\begin{figure}[h]
	\begin{minipage}[b]{0.5\textwidth}
		\centering
		\includegraphics[width=3.3in]{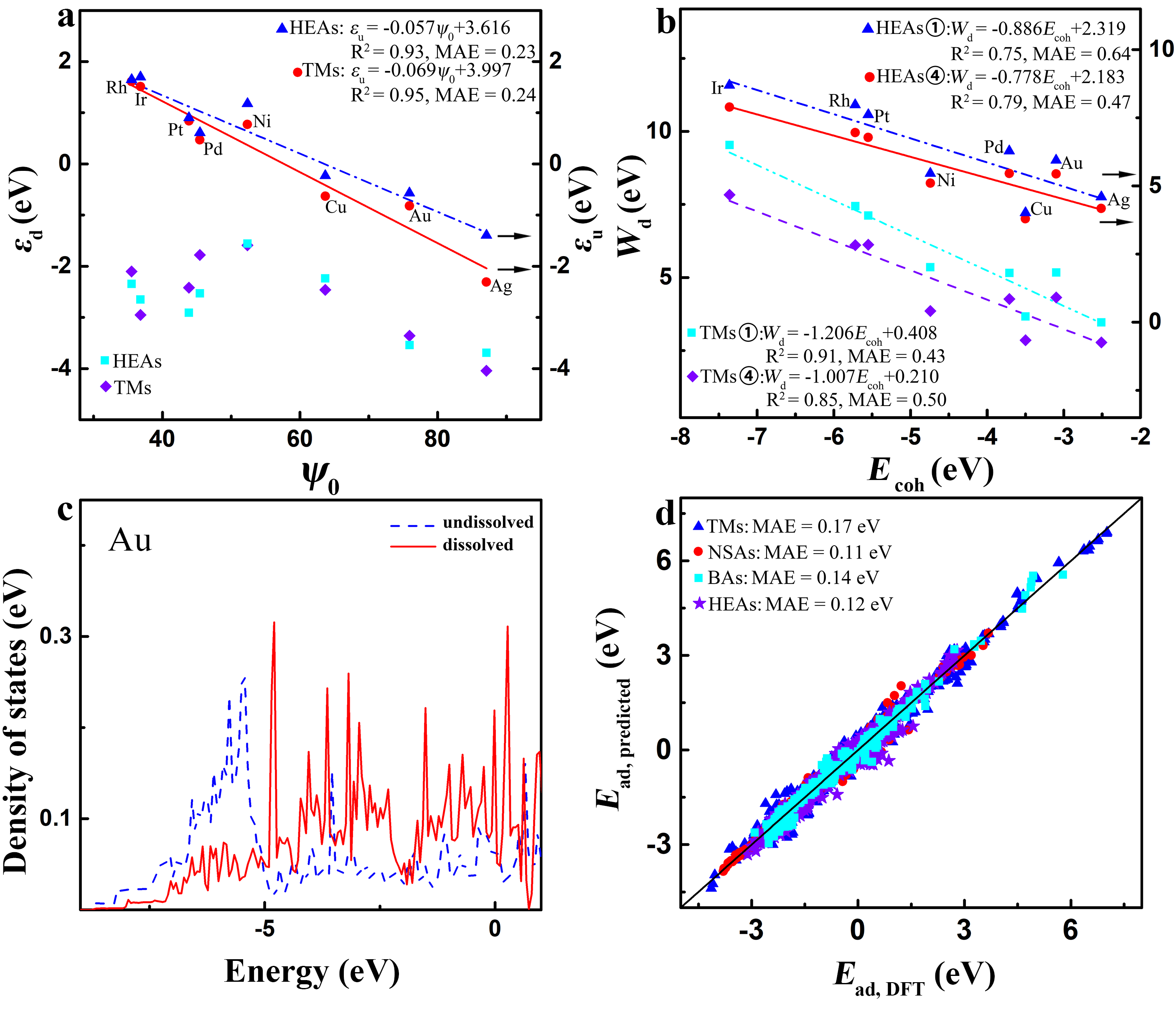}
	\end{minipage}
	\caption{(a) The \textit{d}-band center ($\varepsilon_{\rm d}$) and \textit{d}-band upper edge ($\varepsilon_{\rm u}$) against the electronic descriptor $\psi_0$ for atoms on the undissolved (111) surface of TMs \cite{vojvodic2014electronic} and RuRhIrPdPt-based HEAs. (b) The \textit{d}-band width ($W_{\rm d}$) for surface sites on the undissolved (111) surface (\small{\textcircled{\small{1}}}) and that with three surface atoms dissolved (\small{\textcircled{\small{4}}}), against the cohesive energy \cite{zhang2014understanding}. (c) Density of states of the \textit{s} band of Au atoms on the undissolved (111) surface and that with three atoms dissolved. (d) Comparison between the predicted and DFT-calculated adsorption energies of $\rm {CH}_\textit{x}$ (\textit{x} = 0-3), $\rm {CCH_3}$, CO, $\rm {NH}_\textit{x}$ (\textit{x} = 0-2), OH, F and Cl on TMs, NSAs, BAs, and HEAs \cite{lu2020neural,roling2018structure,andersen2019beyond,hoffmann2016framework,ma2015machine,calle2012physical,xin2012electronic,xin2012predictive,mamun2019high,ghosh2017water,gan2012catalytic}.}
	\label{fm1}
\end{figure}
\noindent where ${\psi_0^2}/\psi_1$ automatically transforms to $\psi_0$ for TMs, implying that our electronic gradient descriptor of adsorption energy is universal for TMs and alloys. Since the $E_{\rm {coh}}$ term corresponds to the effect of adsorption-site atoms, namely the local electronic effect, while the $\psi^{'}$ term considers the coupling between the adsorption site and its surrounding environments, the prefactors $k_1$ and $k_2$ can effectively reflect the electronic localization of alloys in adsorption. We find that $k_1$ = 1/10 and $k_2$ = 9/10 are the optimum values for describing the adsorption-site effect of NSAs, BAs, and HEAs on the adsorption energy. Accordingly, Eqs. (2) and (4) can describe well the adsorption energy of $\rm {CH}_\textit{x}$ (\textit{x} = 1-3), CO, $\rm {NH}_\textit{x}$ (\textit{x} = 1, 2), and OH at the top site of NSAs, BAs, and HEAs (Fig. 2c-e and Figs. S14b, S20 and S21 \cite{li2021submission}), where the fitted MAEs are merely 0.18 eV for NSAs, 0.12 eV for BAs, and 0.14 eV for HEAs. Moreover, the fitted prefactors $\mu_1$ and $\mu_2$ are also consistent with the predictions by Eq. (2) (Tables I, S2 and S3 \cite{li2021submission}). Namely, the prefactors $\mu_1$ and $\mu_2$ of alloys are also determined by the bonding characters $X_{\rm{m}}$ and $X$ as those of TMs do, which comply with the first-order approximation of EMT \cite{norskov1980effective} and bond-order conservation framework, indicating that the homogeneous-electron-gas model, EMT, is suitable for describing the adsorption behavior on both TMs and alloys (Note S4 \cite{li2021submission}).

We now turn to understand the environment effect of alloying on the adsorption energy. Fig. 2c and d and Fig. S22 \cite{li2021submission} show that the adsorption energies of $\rm {CH}_\textit{x}$ (\textit{x} = 0-3), CO, $\rm {NH}_\textit{x}$ (\textit{x} = 0-2), OH, F and Cl at the top site of Ag-based BAs, Pt(Pd)-based NSAs \cite{ma2015machine,calle2012physical,xin2012electronic,xin2012predictive} and Ru(Cu)RhIrPdPt-based HEAs in the environment effect of alloying also follow the linear relations in the adsorption-site effect of alloying with the fitted MAE 0.06 eV for NSAs, BAs, and HEAs, which further demonstrates the robustness of our scheme. Notably, the prefactors $k_1$ and $k_2$ in Eq. (4) are different upon adsorbing on the different elements of HEAs in the environment effect of alloying, indicating the distinct localization of the different alloying elements in determining the adsorption energy. In particular, the localization of RuRhIrPdPt HEAs in adsorbing $\rm {CH}_2$  obeys the order of Ru $>$ Ir $>$ Pt $>$ Pd $>$ Rh (Note S5 and Table S4 \cite{li2021submission}).

When both the adsorption site and surrounding environments change simultaneously from one system to the next, Eqs. (2) and (4) can also capture these complicated alloying effects. For instance, the adsorption of OH at the bridge site of HEAs, the adsorption of CH, CO, C and OH at the top, bridge and hollow sites of BAs and the adsorption of OH, H and OH+H at the hollow site of NSAs comply with our model by Eqs. (2) and (4), with the fitted MAE of 0.05 eV for NSAs, 0.17 eV for BAs, and 0.08 eV for HEAs (Fig. 2f and Figs. S23-S26, Tables S2-S6 and Note S6 \cite{li2021submission}) \cite{lu2020neural,andersen2019beyond,mamun2019high,gan2012catalytic,ghosh2017water}. Overall, our scheme is universal for capturing the adsorption energy on TMs and alloys in dissolution.  

Our scheme reveals a novel physical picture for surface adsorption. The \textit{d}-band model by N$\o$rskov et al. \cite{hammer1995gold,hammer1995electronic,hammer2000theoretical,vojvodic2014electronic}, which takes the \textit{d}-band center or \textit{d}-band upper edge as the descriptors of adsorption and assumes a constant contribution of \textit{s} bands to adsorption for metals, is mainly applied into TMs but not complex alloys like BAs and HEAs \cite{andersen2019beyond}. Our scheme, which is effective in complex alloys, can also be related to the \textit{d}-band and \textit{s}-band properties of TMs but in a significantly different way from the \textit{d}-band model. $\psi^{'}$ reflects the atoms$'$ \textit{d}-band upper-edge ratio between the adsorption site and its nearest-neighbors (Fig. 3a), whereas the adsorption-site atoms$'$ cohesive energy does their own \textit{d}-band width and \textit{s}-band depth \cite{turchanin2008cohesive}. Namely, the \textit{d}-band upper-edge ratio of active centers and the \textit{d}-band width and \textit{s}-band depth of adsorption sites together control the alloying effects in adsorption. The dissolution of a surface atom reduces significantly its neighbors$'$ \textit{d}-band width and \textit{s}-band depth (Fig. 3b and c and Fig. S27 \cite{li2021submission}), making their cohesive energy crucial to determining the adsorption energy on metallics in dissolution. Notably, the exclusion of the \textit{s}-band contribution makes the model lose its prediction ability by increasing the MAE to $>$ 0.32 eV (Fig. S28 and Table S7 \cite{li2021submission}), demonstrating a material-dependent contribution of \textit{s} bands of adsorption sites to the adsorption of complex alloys. Clearly, our scheme goes beyond the \textit{d}-band model in understanding the adsorption on TMs and alloys in dissolution.

Now we access the prediction accuracy of our model. Encouragingly, our model, Eqs. (2-4), is convenient to predict the adsorption energy on metallics as all parameters are easily accessible. The MAE of the predicted adsorption energies relative to the DFT results is $\sim$0.14 eV for the adsorption of $\rm {CH}_\textit{x}$ (\textit{x} = 0-3), $\rm {CCH_3}$, CO, $\rm {NH}_\textit{x}$ (\textit{x} = 0-2), OH, F and Cl on TMs, NSAs, BAs, and HEAs (Fig. 3d) regardless of the adopted (semi-)local functionals (Note S6 \cite{li2021submission}), which is less than the approximate error of (semi-)local functionals, ±0.2 eV. This model is also predictive for the reactivity at surfaces including the activation energy, reaction energy, and experimentally measured activity for 11 different reactions, and even including those on HEAs that can break the BEP relation (Note S7 and Figs. S29-S40 \cite{li2021submission}). Considering the large amount (2586 adsorption energies and 539 reactivity data) and wide window (11.16 eV for adsorption energy and 6.30 eV for reactivity) of data as well as the easy accessibility of the descriptors, our model exhibits promising prediction power for surface adsorption and reaction on TMs and alloys.

In summary, we have identified the electronic gradient and cohesive energy of surface sites as new powerful descriptors for understanding surface adsorption and reaction on metallics in dissolution. With these descriptors, a predictive model has been established for the quantitative determination of the adsorption energy on TMs, NSAs, BAs, and HEAs. Our scheme is a crucial development of \textit{d}-band model, revealing that the \textit{s}-band contribution is material-dependent and indispensable in the accurate description of the alloying and dissolution effects on surface adsorption and reaction. The electronic gradient descriptor ${\psi_0^2}/\psi_1$, corresponding to the \textit{d}-band upper-edge ratio of active centers, is crucial for determining the adsorption energy in both the adsorption-site and environment effects of alloying, whereas the cohesive-energy descriptor, corresponding to the \textit{d}-band width and \textit{s}-band depth of adsorption sites, plays an increasingly important role in determining the adsorption energy as the alloying and dissolution proceed and is dominant in the adsorption-site effect of alloying. This scheme unravels the fundamental physics of surface adsorption on TMs and alloys, and provides a convenient predictive tool for the rational design of advanced alloy catalysts.

We gratefully acknowledge support from the National Natural Science Foundation of China (Nos. 22173034, 21673095, 11974128, 51631004, 52130101), the Opening Project of State Key Laboratory of High Performance Ceramics and Superfine Microstructure (SKL201910SIC), the Program of Innovative Research Team (in Science and Technology) in University of Jilin Province, the Program for JLU (Jilin University) Science and Technology Innovative Research Team (No. 2017TD-09), the Fundamental Research Funds for the Central Universities, and the computing resources of the High Performance Computing Center of Jilin University, China.

\bibliography{paper}

\end{document}